\documentclass[conference]{IEEEtran}
\IEEEoverridecommandlockouts

\usepackage{cite}
\usepackage{amsmath,amssymb,amsfonts}
\usepackage{algorithmic}
\usepackage{graphicx}
\usepackage{textcomp}
\usepackage{xcolor}
\usepackage{cleveref}
\usepackage{multicol, multirow}

\def\BibTeX{{\rm B\kern-.05em{\sc i\kern-.025em b}\kern-.08em
    T\kern-.1667em\lower.7ex\hbox{E}\kern-.125emX}}

\Crefname{equation}{Eq.}{Eqs.}
\Crefname{figure}{Fig.}{Figs.}
\Crefname{section}{Sec.}{Sec.}

\begin{document}
\title{CJST: CTC Compressor based Joint Speech and Text Training for Decoder-Only ASR}

\author{
\IEEEauthorblockN{
Wei Zhou, Junteng Jia, Leda Sari, Jay Mahadeokar, Ozlem Kalinli}
\IEEEauthorblockA{\textit{Meta AI}}
zhouw@meta.com
}

\maketitle

\begin{abstract}
CTC compressor can be an effective approach to integrate audio encoders to decoder-only models, which has gained growing interest for different speech applications.
In this work, we propose a novel CTC compressor based joint speech and text training (CJST) framework for decoder-only ASR.
CJST matches speech and text modalities from both directions by exploring a simple modality adaptor and several features of the CTC compressor, including sequence compression, on-the-fly forced peaky alignment and CTC class embeddings.
Experimental results on the Librispeech and TED-LIUM2 corpora show that the proposed CJST achieves an effective text injection without the need of duration handling, leading to the best performance for both in-domain and cross-domain scenarios.
We also provide a comprehensive study on CTC compressor, covering various compression modes, edge case handling and behavior under both clean and noisy data conditions, which reveals the most robust setting to use CTC compressor for decoder-only models.
\end{abstract}


\section{Introduction \& Related Work}
With the great success of large language models (LLM) \cite{gpt4, llama2, gemini, llama3}, decoder-only architectures have been widely adopted for various speech applications \cite{speechLlama24meta, wu23ctc-comp-do-speechLlama, tusnoo24streamingDecoderOnly}.
To integrate speech into decoder-only models, an audio encoder is commonly used to extract high-level representations.
The extracted acoustic embeddings can be directly fed into the decoder model in the continuous space, or can be discretized first to allow the decoder-only model operate with discrete audio tokens \cite{zhang23speechGPT, AudioPaLM} in the same way as text tokens.
While the latter has become a popular research area recently, this work focus on the continuous representations for the task of automatic speech recognition (ASR).
More specifically, we aim to improve decoder-only ASR models of production-deployable size without using external LMs.

One common way of feeding continuous acoustic embeddings into a decoder-only model is via an adaptor.
A simple and widely-used adaptor is a time reduction layer followed by a linear projection layer \cite{speechLlama24meta, ma24simpleLLM-ASR}.
Another approach is to use connectionist temporal classification (CTC) \cite{graves2016ctc} to compress the sequence length based on CTC probabilities, which is referred to as CTC compressor.
The concept of CTC compressor was originally proposed for speech translation \cite{gaido21ctc-compress} using attention-based encoder-decoder models \cite{bahdanau2016end, chan2016listen}.
An additional CTC layer was introduced in the encoder to compress the sequence length by averaging neighboring frames of the same CTC predictions.
This idea was further explored in \cite{wu23ctc-comp-do-speechLlama} for speech translation using decoder-only models, where the compression can also be done by removing blank frames.
Various extensions of the CTC compressor were also explored, e.g., for fully formatted ASR transcription \cite{ling24ctcLLM} and for streaming ASR \cite{tusnoo24streamingDecoderOnly}.

Joint speech and text training is another important research topic to improve ASR performance, especially when no external LM can be used.
An effective text injection would require certain level of speech and text modality matching.
This can be implicit at the model output level by applying the same objective and target to both speech and text input, such as JOIST \cite{sainath22JOIST} and textogram \cite{thomas22textogram}.
This can also be explicit at the model input level by matching speech-text representations, such as MAESTRO \cite{MAESTRO}. 
All these approaches were proposed for recurrent neural network transducer (RNN-T) \cite{graves2012rnnt} models, aiming to bring the text modality closer to the speech one with simple or sophisticated duration handling.
For decoder-only models, this modality aligning is usually not studied for ASR, e.g., implicit aligning in the discrete space for pretraining \cite{spiritLM, zhang24speechLM} or input space aligning for response matching \cite{AudioChatLlama}.

In this work, we propose a novel CTC compressor based joint speech and text training (CJST) framework for decoder-only ASR.
More specifically, we propose to utilize the CTC compressor to match speech and text representations from both directions.
This is done by combining a simple modality adaptor with several features of the CTC compressor, including sequence compression, on-the-fly forced peaky alignment and CTC class embeddings.
Without the need of duration consideration, the proposed CJST achieves effective text injection for both in-domain and cross-domain scenarios, leading to consistent improvements.
We also provide a comprehensive study of the CTC compressor, covering various compression modes, edge case handling and behavior under both clean and noisy data conditions, which reveals the most robust setting to use CTC compressor for decoder-only models.

\section{Extended CTC Compressor}
\begin{figure}[t]
\begin{minipage}[b]{1.0\linewidth}
  \centering
  \centerline{\includegraphics[width=7cm]{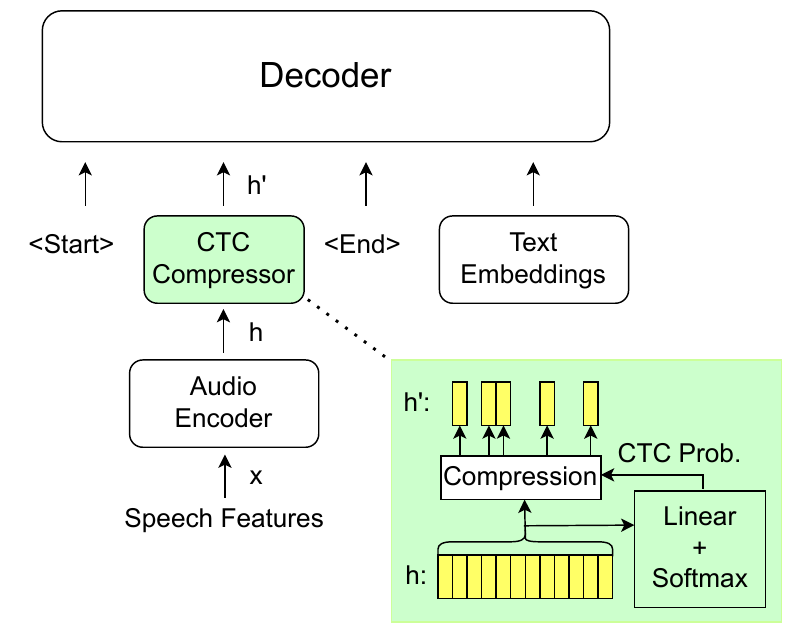}}
\end{minipage}
\vspace{-0.5cm}
\caption{\it Overview of decoder-only model structure and CTC compressor.}
\label{fig:model}
\vspace{-0.4cm}
\end{figure}

An overview of the decoder-only model structure applied in this work can be seen in \Cref{fig:model}.
The {\it \textless Start\textgreater} and {\it \textless End\textgreater} symbols are general placeholders for optional (special) tokens surrounding the acoustic embeddings to form the complete prompt.
The CTC compressor is directly applied at the output of the audio encoder.
All the acoustic embeddings $h$ are passed to a linear layer followed by a softmax to generate CTC probabilities, which are used to compress $h$ under different modes.
The compressed output $h'$ are then fed into the decoder model.
Here an additional linear projection can be applied for dimension matching, which is omitted for simplicity.

\subsection{Compression Modes}
Based on the CTC probabilities, different methods can be applied for the compression shown in \Cref{fig:model}.
In this work, we investigate the following four modes for both clean and noisy data conditions:
\begin{enumerate}
    \item {\bf Blank Prediction Removal}: Based on the greedy CTC predictions, all blank frames are removed \cite{wu23ctc-comp-do-speechLlama, tusnoo24streamingDecoderOnly}.
    \item {\bf Same Prediction Average}: Based on the greedy CTC predictions, neighboring frames of the same predictions are averaged \cite{gaido21ctc-compress, wu23ctc-comp-do-speechLlama}.
    \item {\bf Blank Probability Removal}: All frames with blank probabilities higher than a predefined threshold are removed \cite{ling24ctcLLM}.
    \item {\bf Combine 3 + 2}: We firstly apply blank probability removal and then apply same prediction average to achieve a more compact output.
\end{enumerate}
These approaches cover most existing ones in the literature (and beyond).
All of them largely rely on the quality of the CTC probabilities. 
Therefore, unlike \cite{wu23ctc-comp-do-speechLlama}, we apply the CTC compressor on the top instead of the middle layer of the audio encoder for better CTC modeling.
Additionally, in training, we always include an auxiliary CTC loss in the compressor.

\subsection{Special Case: Empty Output}
\label{sec:empty}
One edge case about CTC compressor is that it may produce empty output by removing all the input frames.
This was not addressed in any previous work.
We observe that this can happen sometimes for low-quality data such as strong/pure noise, music or even wrong languages, where the transcripts are also erroneous.
Here we propose two solutions for this problem:
\begin{itemize}
    \item {\bf Empty Skip}: We skip these utterances in training and let the model directly output EOS in inference. The intuition here is that those low-quality data only do more harm than good in terms of model training.
    
    \item {\bf Empty Fallback}: We average all the encoder output $h$ to a single frame, and then perform training and inference as usual. The intuition here is that some of these data might still be useful to improve the model's robustness.
\end{itemize}

\subsection{Embedding Sharing}
The CTC compressor effectively applies a dynamic time reduction on the encoder output based on the speech content.
Ideally, the compressed acoustic embeddings $h'$ become more aligned with the underlying token sequence in terms of both length and content, which may facilitate the decoder model's learning of speech content discrimination.

This aligning can be further enforced from a representation level by sharing the text embeddings with the CTC class embeddings, i.e., the weights of the linear layer in the compressor.
In this case, the audio encoder output are trained to be closer to the text embeddings by the CTC objective.
This concept was also raised in \cite{ling24ctcLLM} without showing experimental effects.
In this work, we also investigate whether such enforced link between speech and text representations is helpful for ASR.

\begin{figure}[t]
\begin{minipage}[b]{1.0\linewidth}
  \centering
  \centerline{\includegraphics[width=7cm]{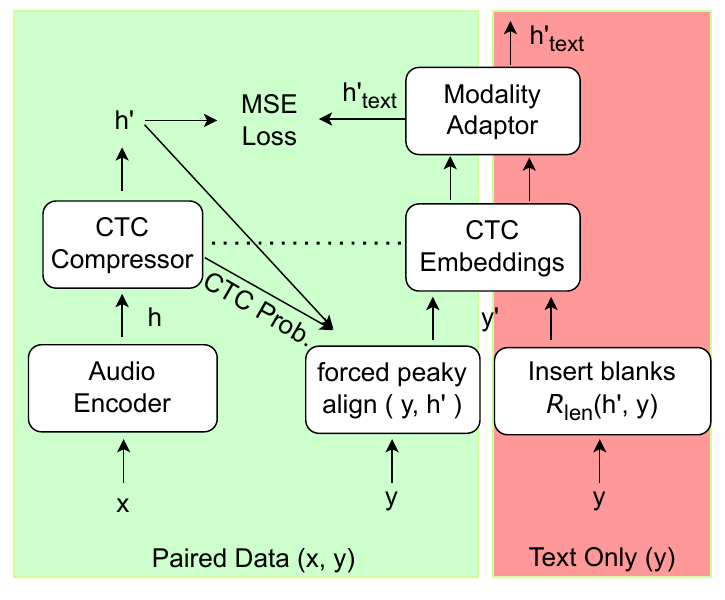}}
\end{minipage}
\vspace{-0.8cm}
\caption{\it Proposed CJST framework}
\label{fig:cjst}
\vspace{-0.4cm}
\end{figure}

\section{Joint Speech and Text Training}
For joint speech and text training, we mainly consider paired speech-text and text-only data in this work.
Here an effective training on large amounts of text-only data is the key factor.

\subsection{LM-like}
As decoder-only models are naturally language models (LM), the most straightforward way to perform text injection is to train the model on text-only data as a LM, i.e., no audio prompt at all.
In the context of ASR, this can be viewed as internal LM (ILM) training \cite{variani2020hat, Meng2021ILMT} for decoder-only models.
We show that this LM-like training is already quite effective, which is used as a strong baseline for comparison.

\subsection{Proposed CJST}
In all previous work on the CTC compressor, it is mainly used to bring the speech modality closer to the text modality, especially on the sequence length perspective.
While this is still true in this work, we propose to further utilize the CTC compressor for text-to-speech representation matching without the need of duration consideration.
This yields the proposed CJST for joint speech and text training, which further explores two unused features of the CTC compressor, namely, on-the-fly forced peaky alignment and CTC class embeddings.
The main idea of CJST is summarized in \Cref{fig:cjst}.

For {\bf paired speech-text data}, we forward the speech features $x$ through the model and perform ASR training as usual.
Additionally, we take the compressed acoustic embeddings $h'$ together with their CTC probabilities to perform forced peaky alignment with the ground truth text tokens $y$ by disallowing label loops.
In this case, the number of possible alignment sequences is rather small due to sequence compression and no loop transitions. 
Therefore, this step can be applied on the fly with only minor cost on the training throughput. 
The obtained forced alignment $y'$ is effectively extending $y$ to the length of $h'$ by inserting blanks.
We then pass $y'$ to the CTC embeddings of the compressor and an modality adaptor to generate pseudo acoustic embeddings $h'_{\text {text}}$.
A mean square error (MSE) loss is applied between $(h', h'_{\text {text}})$ to train the modality adaptor solely. 
Additionally, we also record the length ratio between $h'$ and $y$ in a moving average fashion, denoted as $R_{\text{len}}(h', y)$.

For {\bf text-only data}, we randomly insert blanks into $y$ based on the recorded length ratio $R_{\text{len}}(h', y)$.
Specifically, for each token in $y$, one blank symbol can be inserted beside it under the probability of $\max (0, R_{\text{len}}(h', y) - 1)$.
This leads to a simulated alignment sequence $y'$ as in the paired-data case.
Similarly, the $y'$ is passed through CTC embeddings and the modality adaptor to generate $h'_{\text {text}}$.
Finally, this text-based pseudo acoustic prompt is used to train the decoder model with the ASR objective as usual.
This makes the training more consistent with inference where an acoustic prompt is always present.
Note that we stop the gradient at $h'_{\text {text}}$ for text-only training.
Additionally, we randomly mask 20\% elements of $h'_{\text {text}}$ to 0 to maintain the learning difficulty.

\section{Experiments}
\subsection{Data}
We perform experiments on the 960-hour Librispeech dataset \cite{libsp} and 2M-hour in-house data.
Our in-house data is transcribed from de-identified video data of varying acoustic conditions, which is further augmented with speed perturbation \cite{SpeedPerturb}, simulated reverberation and randomly sampled additive background noise extracted from videos.
The in-house data contains audios of varying length up to 300 seconds.
We randomly select 1k and 3k utterances from it as dev and test sets, which are about 14 and 42 hours, respectively.
For joint speech and text training, we use the Librispeech and TED-LIUM2 \cite{tedlium2} LM text data for in-domain and out-of-domain scenarios.

\begin{table}[t]
\caption{\it WER [\%] results of various encoder integration into the decoder-only model on Librispeech test sets. Different encoder input $\times$ output time reduction is applied for the adaptor approach, while the CTC compressor uses a fixed encoder input time reduction of 6.}
\begin{center}
\begin{tabular}{|l|c|c|c|}
\hline
Encoder Integration & test-clean & test-other & $R_{\text{len}}(h', y)$ \\ \hline
{\bf Adaptor (stacking + linear)} & &&\\
$4 \times 5$ & 3.33 & 5.50 & 1.47 \\
$4 \times 6$ & 3.41 & 5.54 & 1.22  \\
$6 \times 3$ & 3.28 & 5.58 & 1.64 \\
$6 \times 4$ & 3.38 & 5.63 & 1.22 \\
$8 \times 3$ & 3.30 & 5.76 & 1.22 \\ \hline \hline

{\bf CTC Compressor } & && \\
1. blank prediction removal & 2.44 & 5.32 & 1.12 \\ 
\quad \quad + emb-sharing & 2.50 & 5.14 & 1.12 \\ \hline
2. same prediction average & 2.54 & 5.28 & 1.94 \\
\quad \quad + emb-sharing & 2.34 & 5.28 & 1.94 \\ \hline
3. blank probability removal & &&\\
\quad threshold = 0.6 & 2.48 & 5.15 & 1.13 \\
\quad threshold = 0.9 & 2.22 & 5.12 & 1.19 \\
\quad \quad + emb-sharing & 2.19 & 4.98 & 1.19 \\
\quad threshold = 0.95 & 2.22 & 4.94 & 1.22 \\
\quad \quad + emb-sharing & 2.17 & 4.94 & 1.22 \\
\quad \quad + move to encoder middle & 2.37 & 5.20 & 1.22 \\ \hline
4. combine 3 (0.95) + 2 & 2.27 & 5.04 & 1.08 \\
\quad \quad + emb-sharing & 2.23 & 4.92 & 1.08 \\
\hline
\end{tabular}
\label{tab:libsp-ctc}
\end{center}
\vspace{-0.4cm}
\end{table}

\begin{table}[t]
\caption{\it WER [\%] results on the in-house data for the same comparison as done in \Cref{tab:libsp-ctc}. The CTC compressor uses a fixed encoder input time reduction of 8 and applies the empty fallback scheme by default.}
\begin{center}
\begin{tabular}{|l|c|c|}
\hline
Encoder Integration & test & $R_{\text{len}}(h', y)$ \\ \hline
{\bf Adaptor (stacking + linear)} & &\\
$6 \times 4$ & 13.12 & 1.78 \\
$8 \times 3$ & 13.13 & 1.78 \\ \hline \hline

{\bf CTC Compressor } & & \\
1. blank prediction removal & 18.85 & 1.04 \\ 
\quad \quad + emb-sharing & 19.81 & 1.04 \\ \hline
2. same prediction average & 15.56 & 1.9 \\
\quad \quad + emb-sharing & 14.77 & 1.9 \\ \hline
3. blank probability removal &  & \\
\quad threshold = 0.9 & 13.66 & 1.24 \\
\quad \quad + emb-sharing & 13.75 & 1.24 \\
\quad threshold = 0.95 & \bf 12.85 & 1.35 \\
\quad \quad + emb-sharing & 12.87 & 1.35 \\
\quad \quad + empty skip instead of fallback & 13.10 & 1.35\\ \hline
4. combine 3 (0.95) + 2 + emb-sharing & 15.46 & 1.18 \\
\hline
\end{tabular}
\label{tab:video-ctc}
\end{center}
\vspace{-0.5cm}
\end{table}

\subsection{Experiment Settings}
Our decoder model contains 12 LLaMA decoder layers \cite{llama2} with 768 hidden dimension, 12 attention heads and 2048 MLP dimension.
The rotary position embedding \cite{RoPE} has a maximum length of 4096.
We use 4k SentencePiece model (SPM) units \cite{SPM} for each dataset individually.
For the audio encoder, we use 24 conformer layers \cite{Gulati20conformer} with 512 hidden dimension and 8 attention heads.
For each conformer block, we skip the first feed-forward module and swap the convolution and multi-head self attention modules.
An initial VGG network \cite{simonyan2015vgg} is used at the encoder input for time reduction.

For both datasets, we pretrain the audio encoder with CTC loss for 200k steps.
We then train the full model (\Cref{fig:model}) with cross-entropy (CE) loss, where an auxiliary CTC loss of weight 0.5 is applied to the CTC compressor.
The CE training is performed for 200k steps for Librispeech and 500k steps for the in-house data.
SpecAugment \cite{SpecAugment} is always applied.
For joint speech and text training, both speech- and text-based CE losses have a weight 1.0.
For CJST, we use a simple conformer layer for the modality adaptor, which contains single attention head, convolution kernel 3 and no dimension uplifting for the feed-forward module.
It is from-scratch trained with the MSE loss of weight 1.0.
We initialize $R_{\text{len}}(h', y)$ with 1.0.

For recognition, no external LM is used and a simple beam search with beam size 4 is applied.
We use the dev sets to select the best checkpoint and report word error rate (WER) on the test sets.

\begin{table}[t!]
\caption{\it WER [\%] results of from-scratch joint speech and text training on the Librispeech corpus. The baseline models without text injection are from \Cref{tab:libsp-ctc}.}
\begin{center}
\begin{tabular}{|c|c|c|c|}
\hline
Encoder Integration & Text Injection & test-clean & test-other \\ \hline
\multirow{2}{*}{\shortstack[c]{Adaptor ($6 \times 4$)}}& no & 3.38 & 5.63 \\ 
                     & LM-like & 2.54 & 5.26 \\ \hline
\multirow{3}{*}{\shortstack[c]{CTC Compressor\\ (blank 0.95 removal)}} & no & 2.22 & 4.94 \\ 
                               & LM-like & 2.13 & 4.85 \\
                    & CJST & \bf 2.09 & \bf 4.71 \\
\hline
\end{tabular}
\label{tab:jst-scratch}
\end{center}
\vspace{-0.4cm}
\end{table}

\subsection{Comprehensive Evaluation of CTC Compressor }
We firstly perform a comprehensive study on the CTC compressor for both clean and noisy data.
For comparison, we also include the common adaptor approach for encoder integration to decoder-only models, which adopts frame stacking followed by a linear projection.
Various combinations of encoder input time reduction (VGG network) and encoder output time reduction (stacking) are evaluated for the adaptor approach.
The CTC compressor uses a fixed reduction factor at the encoder input.
Besides WER, we also show the average length ratio $R_{\text{len}}(h', y)$ recorded in training to give an idea of sequence compression effect.

The results on the clean Librispeech and the noisy in-house data are shown in \Cref{tab:libsp-ctc} and \Cref{tab:video-ctc}, respectively.
Although all CTC compressor results outperform the adaptor ones on Librispeech, most of them are much worse on the in-house data, especially those greedy prediction based.
This indicates the strong sensitivity of the CTC compressor approach.
The most robust compression mode appears to be the blank probability removal with a high threshold (0.95 here), which achieves best results on both datasets as well as a rather compact sequence length.
Moving the compressor to the encoder middle layer degrades the performance.
The empty output issue described in \Cref{sec:empty} only occurs for the in-house data, where the empty-fallback scheme seems to be a better choice than the empty-skip one.
Additionally, sharing CTC and text embeddings does not show consistent advantages.

We use the blank probability removal with threshold 0.95 for all further experiments.

\subsection{Joint Speech and Text Training}
We evaluate two scenarios for the joint speech and text training.
The first one is from-scratch joint training with 50\% paired speech-text and 50\% in-domain text-only data for each update.
The goal is to verify the training stability as well as to improve the model's in-domain performance.
Note that the audio encoder is still pretrained first.
We perform this experiment with the Librispeech audio and LM training data.
For comparison, we also include one model with adaptor-based encoder integration from \Cref{tab:libsp-ctc}, where text injection can be applied with the LM-like approach.
The results are shown in \Cref{tab:jst-scratch}.
The LM-like text injection can already bring noticeable improvement, especially over the weaker baseline, while the proposed CJST leads to the best results.

The second experiment set aims to simulate real-world scenarios.
Given a seed model trained with sufficient paired speech-text data, we continue training it by injecting both in-domain and out-of-domain text-only data.
The goal is to maintain the same (or even better) in-domain performance while achieving a performance boost on the cross-domain evaluation.
In this case, we keep 20\% paired speech-text data, 30\% in-domain text-only data and 50\% out-of-domain text-only data for the training.
This is performed with the Librispeech audio and LM training data as well as the TED-LIUM2 LM text data, correspondingly.
We continue training the Librispeech base models from \Cref{tab:libsp-ctc} for another 60k steps.
To verify the improvements are not from longer training, we also include results of continuing training the baseline with paired speech-text data.
The WER results on both Librispeech and TED-LIUM2 test sets are shown in \Cref{tab:jst-cont}.
Again, we see that the LM-like text injection is already quite effective, while our proposed CJST consistently achieves further improvements in all cases, fulfilling the goal of this experiment.

Comparing to the baseline trained on paired data only, CJST achieves about 6\% relative improvements for both in-domain (\Cref{tab:jst-scratch}) and cross-domain (\Cref{tab:jst-cont}) scenarios.

\begin{table}[t!]
\caption{\it WER [\%] results of continuing training the Librispeech base models from \Cref{tab:libsp-ctc} with joint speech and text training. Both in-domain Librispeech and out-of-domain TED-LIUM2 (TL2) text-only data are used for text injection.}
\begin{center}
\begin{tabular}{|l|c|c|c|c|c|}
\hline
\multirow{3}{*}{Encoder Integration} & \multirow{3}{*}{\shortstack[c]{Text\\Injection}} & \multirow{3}{*}{\shortstack[c]{Freeze\\Encoder}} & \multicolumn{2}{|c|}{Librispeech} & TL2 \\ 
 &  &  & test & test & \multirow{2}{*}{test} \\ 
 &  &  & clean & other &  \\ \hline
Adaptor ($6 \times 4$) & no & no & 3.38 & 5.63 & 11.45 \\ 
+ train: speech-only & no & no & 3.35 & 5.60 & 11.38 \\ 
+ train: speech + text & LM-like & yes & 3.16 & 5.55 & 10.88 \\ \hline 
CTC Compressor & \multirow{2}{*}{no} & \multirow{2}{*}{no} & 2.22 & 4.94 & 10.75 \\
+ train: speech-only & & & 2.23 & 4.92 & 10.79 \\ \hline
\multirow{4}{*}{+ train: speech + text} & \multirow{2}{*}{LM-like} & yes & 2.22 & 4.88 & 10.32 \\
                        &       & no & 2.22 & 4.83 & 10.32 \\ \cline{2-6}
                         & \multirow{2}{*}{CJST}  & yes & \bf 2.17 & \bf 4.77 & 10.20 \\
                         &       & no & \bf 2.17 & 4.80 & \bf 10.14 \\

\hline
\end{tabular}
\label{tab:jst-cont}
\end{center}
\vspace{-0.5cm}
\end{table}

\section{Conclusion}
In this work, we presented CJST, a novel CTC compressor-based joint speech and text training framework for decoder-only ASR.
With a comprehensive evaluation, we firstly showed that CTC compressor can be sensitive under certain compression modes, while the blank probability removal with a high threshold robustly gives the best performance for both clean and noisy data.
We then showed that LM-like text injection is already quite effective for decoder-only models, while our proposed CJST consistently achieves further improvements by matching speech and text modalities from both directions.
Experiments on both Librispeech and TED-LIUM2 corpora show that CJST achieves the best performance for both in-domain and cross-domain text injection.

\section{Acknowledgment}
\vspace{-0.5mm}
\small
We thank Yashesh Gaur, Yingyi Ma, Zhe Liu, Suyoun Kim and Ke Li for useful discussions.

\bibliographystyle{IEEEtran}
\bibliography{refs}

\end{document}